\documentclass[aps,prl,preprint,groupedaddress,floatfix]{revtex4}
\usepackage{graphicx,amssymb}
\begin{document}

\title{Can ordered precursors promote the nucleation of solid solutions?} 
\author{Caroline Desgranges}
\affiliation{Department of Chemistry, New York University, New York, New York 10003, United States}
\affiliation{Department of Chemistry, University of North Dakota, Grand Forks North Dakota 58202, United States}
\author{Jerome Delhommelle}
\affiliation{Department of Chemistry, New York University, New York, New York 10003, United States}
\affiliation{Department of Chemistry, University of North Dakota, Grand Forks North Dakota 58202, United States}
\date{\today}

\begin{abstract}
Crystallization often proceeds through successive stages that lead to a gradual increase in organization. Using molecular simulation, we determine the nucleation pathway for solid solutions of copper and gold. We identify a new nucleation mechanism (liquid$\to$$L1_2$~precursor$\to$solid solution), involving a chemically ordered intermediate that is more organized than the end product. This nucleation pathway arises from the low formation energy of $L1_2$ clusters which, in turn, promote crystal nucleation. We also show that this mechanism is composition-dependent since the high formation energy of other ordered phases precludes them from acting as precursors.
\end{abstract}

\maketitle

The emergence of order from disorder has long fascinated scientists.\cite{bernstein2002polymorphism,sosso2016crystal} For instance, crystal nucleation, {\it i. e.} formation of a crystal nucleus from a disordered liquid, has attracted considerable attention since the end of the $19^{th}$ century and remains to this day an extremely challenging phenomenon. As acknowledged in Ostwald's famous step rule,\cite{ostwald1897studien} crystallization can be a multi-step process, resulting in a gradual onset of order until the stable crystal polymorph is obtained. There has been tremendous progress in our understanding of this process in recent years. For instance, density functional theory and computer simulations have shed light on the gradual progress of the disordered liquid phase towards the stable crystal phase. In such systems, density fluctuations in the liquid give rise to a crystal nucleus of a metastable form, whose features (density and order) are intermediate between the liquid and the stable polymorph, followed by the formation of the stable polymorph.\cite{oxtoby1998nucleation,tenWolde,desgranges2007controlling} Examples of this progressive onset of order include the pre-organization of liquid pockets, with a structural order intermediate between liquid and crystal,\cite{lechner2011role,desgranges2011role} before the formation of a crystal nucleus, as well as the occurrence of phase separation and demixing prior to crystallization, as observed for polymer mixtures\cite{tanaka1985new} and metal alloys.\cite{desgranges2014unraveling} Here we focus on the crystal nucleation of solid solutions of metals and identify a new nucleation pathway that does not involve the expected gradual increase of order during crystal nucleation. On the contrary, we find that the nucleation pathway can involve a more ordered intermediate structure, since it exhibits both structural and chemical order, than the solid solution, which only has structural order.
 
The formation and properties of solid solutions have attracted considerable attention in recent years.\cite{zhang2017local} For instance, solid solutions of semiconductors, such as $GaN$ and $ZnO$, are remarkably stable and efficient photocatalysts during water splitting.\cite{maeda2005gan} Similarly, solid solutions of $MoS_2$ and $P$ are highly efficient and cost-effective materials for hydrogen production.\cite{ye2016high} Solid solutions of metals also play a key role in the development of multi-principal elements alloys, including high entropy alloys with a superior mechanical performance.\cite{senkov2015accelerated,wu2016thermal} At the other end of the spectrum, metal alloys can also form highly ordered structures with both structural and chemical orders as in the $L1_2$ and $L1_0$ phases, with unique magnetic properties as, for instance, for $FePt$ nanoparticles.\cite{sun2000monodisperse} Understanding the interplay between structural and chemical order is thus key for a wide range of applications.

We focus here on the nucleation process for different compositions of the copper-gold alloy. These systems are characterized by a negative enthalpy of mixing and exhibit solid solutions and ordered phases over a wide composition range, with an order-disorder transition taking place beyond a specific temperature, as shown in the pioneering work of Shockley~\cite{shockley1938theory}. Specifically, we examine this process for a mole fraction of $x_{Cu}=0.25$ ($CuAu_3$ alloy), $x_{Cu}=0.5$ ($CuAu$ alloy) and $x_{Cu}=0.75$ ($Cu_3Au$ alloy).  $CuAu$ has an ordered  $L1_0$ phase, stable below $683$K, and undergoes an order-disorder transition into the disordered close-packed (CP) structure above this temperature. Similarly, $CuAu_3$ and $Cu_3Au$ have an ordered $L1_2$ phase and the disordered CP structure becomes stable above $663$K and $500$K, respectively.\cite{ozolicnvs1998first}

We use molecular simulation to study the nucleation process in $CuAu$ alloys. To model the interactions between metal atoms, we employ the many-body quantum corrected Sutton-Chen embedded atoms model (qSC-EAM)\cite{Luo}. The qSC-EAM force field performs very well when compared to the experiment for the lattice parameters, cohesive energies, surface energies as well as melting points,\cite{kart2005thermal,Supp1} which are key parameters for the crystallization process. Simulations of the nucleation process are carried out in the isothermal-isobaric ensemble within a hybrid Monte Carlo-Molecular Dynamics framework as in previous work on bimetallic systems\cite{desgranges2018unusual,desgranges2014unraveling,desgranges2016effect} at $1$~bar and a temperature 45\% below the melting point obtained for each composition of the alloy.\cite{kart2005thermal} The temperature is set to $754$K for $CuAu_3$, $748$K for $CuAu$ and $737$K for $Cu_3Au$, {\it i.e.} above the order-disorder transition temperatures for all systems. During crystal nucleation, the system has to overcome a large free energy barrier, that corresponds to the formation of a nucleus of a critical size.\cite{tenWolde} Non-Boltzmann sampling methods thus need to be employed to sample the entire nucleation pathway and gain access to the free energy barrier of nucleation. Here we use the umbrella sampling (US) technique,\cite{Torrie} in which the sampling of configurations with low Boltzmann weights is bolstered by a harmonic function of the order parameter\cite{Steinhardt} $Q_6$. The $Q_6$ order parameter measures the rate of crystallinity as shown in prior experimental\cite{Gasser} and simulation studies.\cite{tenWolde,desgranges2007controlling} This approach has been shown to be very efficient for nucleation in single-component systems\cite{tenWolde,Auer,JACS3,Beckham2}, as well as in binary systems and bimetallic alloys.\cite{valeriani2005rate,desgranges2018unusual,desgranges2014unraveling} We add that Frenkel, Dijkstra and co-workers have carried out thorough comparisons of several simulations methods for crystal nucleation.\cite{valeriani2005rate,filion2010crystal} Specifically, Dijkstra and co-workers performed US simulations to determine the free energy barrier, nucleation rate, size and structure of the crystal nucleus. They showed that the US results were in excellent agreement with those obtained using other approaches, such as forward-flux sampling and molecular dynamics simulations.\cite{filion2010crystal} Similar conclusions were drawn by Frenkel and co-workers on crystal nucleation in molten sodium chloride.\cite{valeriani2005rate}

\begin{figure}
\begin{center}
\includegraphics*[width=7cm]{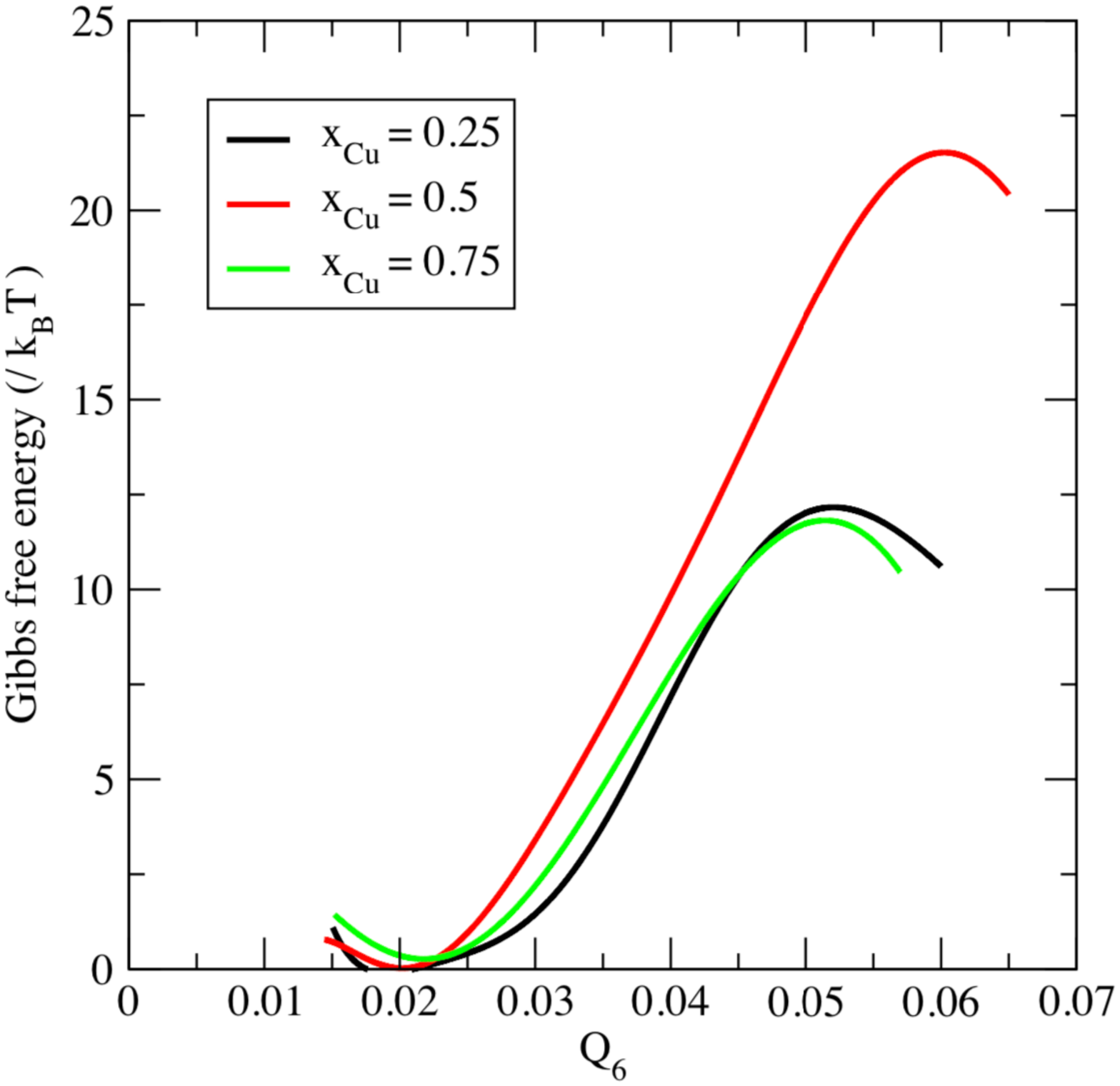}(a)
\includegraphics*[width=7cm]{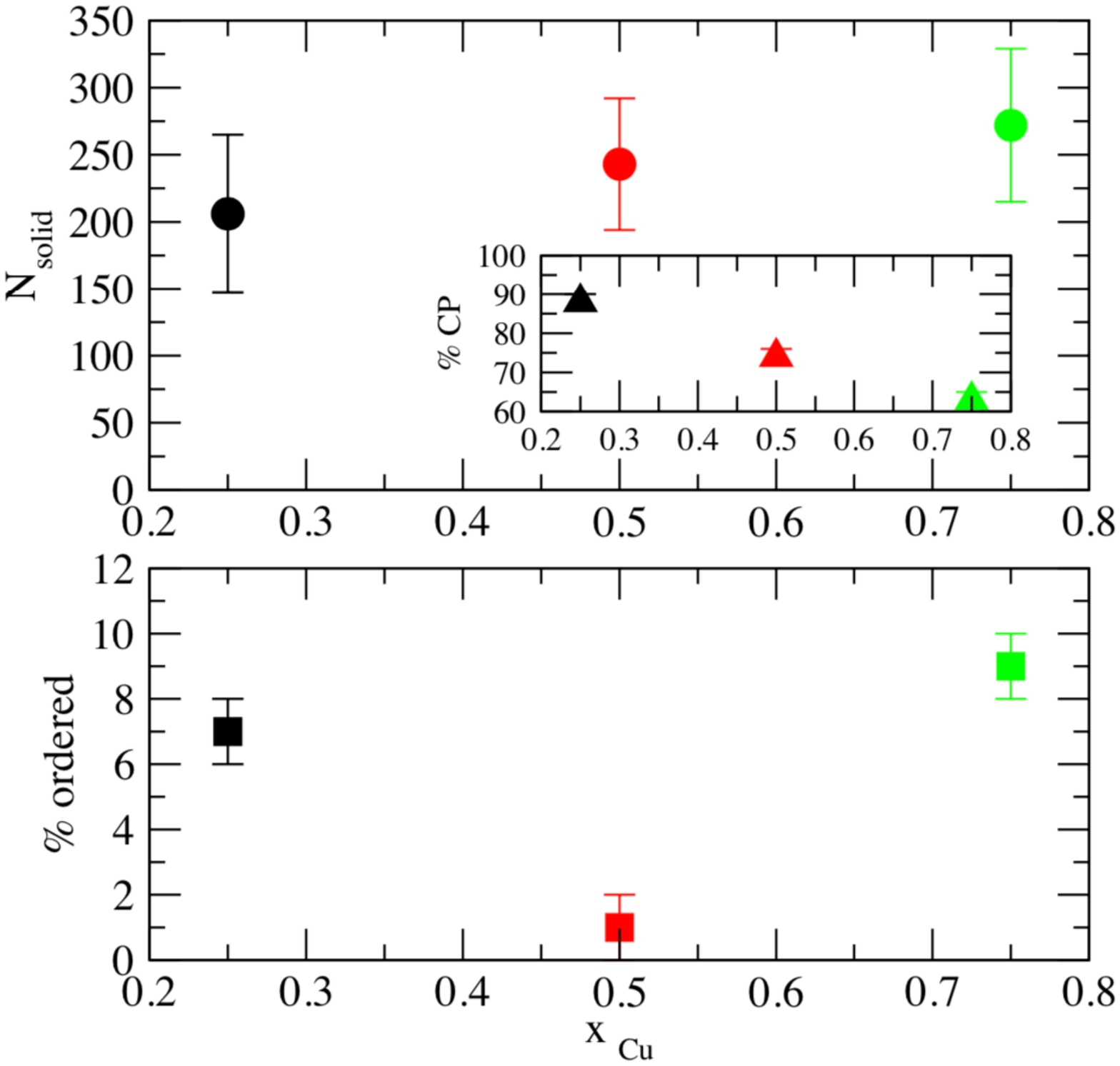}(b)
\end{center}
\caption{Crystal nucleation in $CuAu$ alloys. (a) Free energy barrier of nucleation. (b) (top) Size and structural (inset plot) features for the critical nucleus and (bottom) fraction of ordered structure in the critical nucleus ($L1_0$ for $x_{Cu}=0.5$ or $L1_2$ for $x_{Cu}=0.25$ and $x_{Cu}=0.75$).}
\label{Fig1}
\end{figure}

We start by presenting the results for the free energy barrier of nucleation. Fig.~\ref{Fig1}(a) shows the free energy profiles as a function of the order parameter. The profiles are qualitatively the same for the three mole fractions. First, the free energy profile starts from a minimum, corresponding to the metastable liquid phase, which is disordered and associated with a low value of the order parameter. Then, as the order parameter increases and the crystal nucleus forms, the free energy increases until it reaches a maximum, corresponding to the formation of a critical nucleus. There is, however, a striking difference between the results obtained for $x_{Cu}=0.5$ and the other two mole fractions. The height of the free energy barrier for $CuAu$ is $21.5\pm2~k_BT$, almost twice that for the other two systems. Moreover, the free energy barriers for $CuAu_3$ and $Cu_3Au$ are roughly the same ($12.1\pm1~k_BT$ and $11.8\pm1~k_BT$, respectively). This suggests that the nucleation mechanism in $CuAu$ is very different from the other two systems, and that nucleation proceeds in a similar fashion for $Cu_3Au$ and $CuAu_3$. 

To identify the reason for these different behaviors, we examine the features of the system at the end of the nucleation step. We carry out a detailed analysis of the size and structure of the critical nucleus as in prior work.\cite{Auer,desgranges2018unusual,desgranges2014unraveling} The size of the critical nucleus is plotted, for each composition, in Fig.~\ref{Fig1}(b). This graph shows that the size of the critical nucleus increases steadily with the mole fraction in $Cu$. This increase in size is correlated with the steady decrease in the free energy difference between the solid and liquid $\Delta \mu$ that drives the nucleation process (numerical data are provided in the supplemental information). This is, however, a monotonic trend that cannot account for the unexpectedly high free energy for $x_{Cu}=0.5$, or for the similar results obtained for $x_{Cu}=0.25$ and $x_{Cu}=0.75$. We also show the structure of the critical nucleus in Fig.~\ref{Fig1}(b) as a function of $x_{Cu}$, with a steady decrease in the fraction of atoms with a close-packed-like (CP-like) environment with $x_{Cu}$. Again, we find a monotonic trend as a function of $x_{Cu}$ that does not provide a rationale for the free energy results. We finally recall that crystal nucleation is carried out here above the order-disorder transition for all compositions. As shown at the bottom of Fig.~\ref{Fig1}(b), the fraction of ordered structures, be it $L1_2$ or $L1_0$, is very small (less than $10$\%) in the critical nucleus for all $x_{Cu}$. However, we do observe a non-monotonic dependence of this fraction as a function of $x_{Cu}$, which is a first clue that suggests that ordered structures play a role in the behavior observed for the free energy.

 A closer inspection of the free energy profiles of Fig.~\ref{Fig1}(a) reveals that, even for low values of $Q_6$, the slope of the free energy profile is much steeper for $x_{Cu}=0.5$. This suggests a different nucleation mechanism in the early stages of the process. It has been known since the pioneering work of Franck\cite{Franck} that locally favored structures of low energy, such as icosahedra, form in the supercooled liquid and can impact nucleation. Kawasaki and Tanaka\cite{kawasaki2010formation} have recently confirmed that the supercooled liquid is spanned by transient structures, including medium-range crystal-like domains, that can promote crystal nucleation. Can the presence of different types of low energy precursors in the supercooled liquids account for the different nucleation mechanisms?

 \begin{figure}
\begin{center}
\includegraphics*[width=7.3cm]{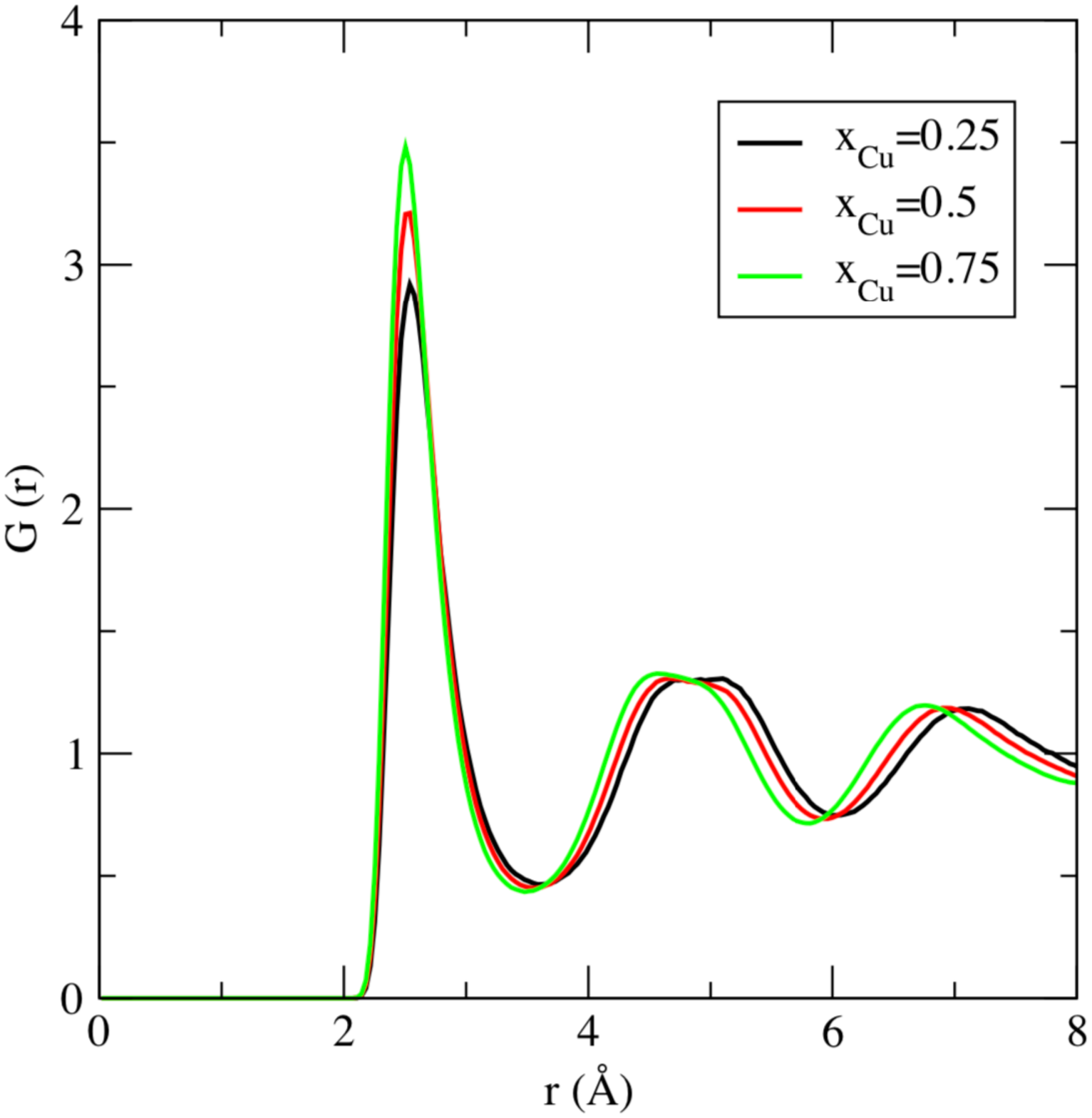}(a)
\includegraphics*[width=7.6cm]{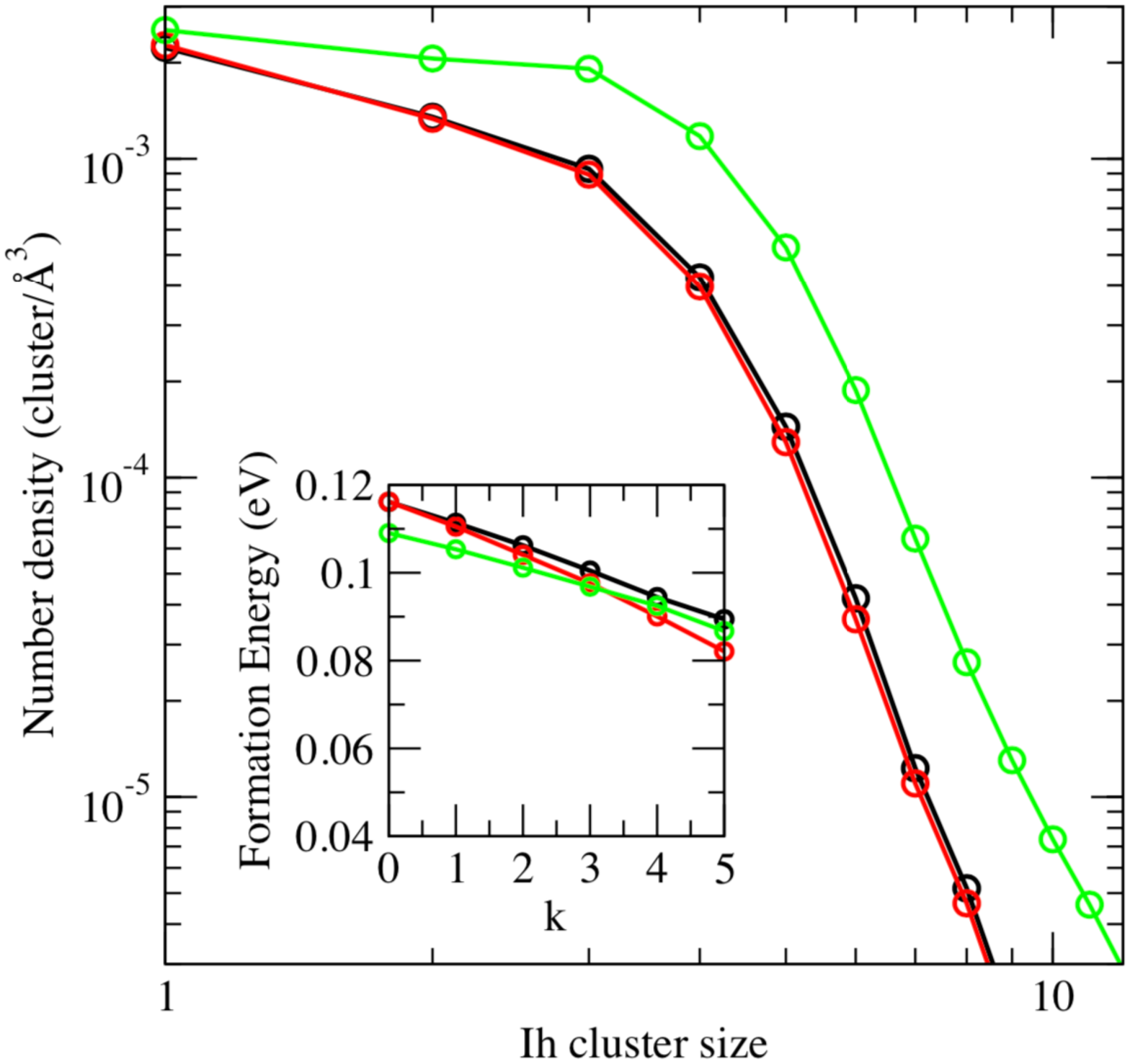}(b)
\end{center}
\caption{Supercooled liquids for the copper-gold alloy. (a) $Cu-Cu$ pair correlation functions. (b) Number density distributions of clusters formed by connected ISROs via volume sharing, with (inset plot) ISRO formation energy as a function of the node degree $k$}
\label{Fig2}
\end{figure}

We start with the characterization of icosahedral order in the supercooled liquids. Icosahedral order is often found in supercooled metallic liquids,\cite{liang2014influence} most notably when there is a significant size difference as here between Au and Cu with $r_{Au}:r_{Cu}=1.13$. The radial distribution functions in Fig.~\ref{Fig2}(a) reveal the onset of splitting for the second peak for the three supercooled liquids. Such a behavior has been linked to the existence of icosahedral order.\cite{liang2014influence} Since icosahedral order is not structurally compatible with crystalline order, the presence of icosahedral clusters (icosahedral short-range order or ISRO) and of connected icosahedra (medium-range order or MRO) is an inhibitor of crystal nucleation\cite{desgranges2018unusual} and often leads to the onset of glassy behavior.\cite{leocmach2012roles} We therefore test the following hypothesis: is the increase in height of the free energy barrier for $x_{Cu}=0.5$ caused by icosahedral order? We determine the number density of clusters of connected icosahedra as a function of the node degree $k$ following the analysis carried out in prior work.\cite{wu2016critical,desgranges2018unusual,desgranges2007molecular} We also calculate the formation energy of these clusters as a function of $k$, choosing as the reference for the energy of each atom the value for crystals of pure $Cu$ and $Au$ in line with Wu {\it et al.}\cite{wu2016critical} Results are shown in Fig.~\ref{Fig2}(b) and confirm the presence of icoasahedral MRO. We also observe that the $x_{Cu}=0.5$ supercooled liquid exhibits the same qualitative features as for the other two mole fractions. The number density plot for connected icosahedra are very similar for $x_{Cu}=0.25$ and $x_{Cu}=0.5$, while the number density is slightly greater for $x_{Cu}=0.75$ as a result of the lower formation energy especially at low $k$ values. However, the formation energy varies very slowly as a function of $k$ for all three systems. For instance, for $x_{Cu}=0.5$, it decreases from $0.12eV$ ($k=0$) to $0.08eV$ ($k=5$) and never reaches a level that can actually impact nucleation. Previous work on supercooled liquids of $CuZr$ and $Ag_6Cu_4$ have reported 2- to 4-fold decreases in the formation energy~\cite{wu2016critical} over the same range of $k$, and values of the order of $0.05eV$ for $k=5$ for the development of MRO structures large enough to impact the height of the free energy barrier of nucleation.~\cite{desgranges2018unusual} We thus conclude that the presence of icosahedral SRO and MRO does not account for the different nucleation mechanisms observed here.  

\begin{figure}
\begin{center}
\includegraphics*[width=7.5cm]{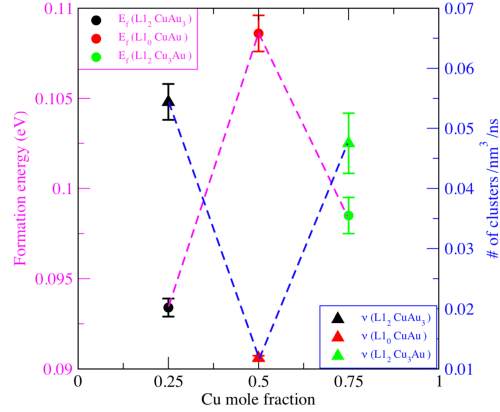}
\end{center}
\caption{Formation energy (left y-axis) of ordered $L1_2$ and $L1_0$ clusters, together with the number of ordered clusters (right y-axis) per unit volume and unit time.}
\label{Fig3}
\end{figure}

Given the ability of copper-gold alloys to form ordered structures, we now look for the possible presence of clusters with a $L1_0$ or a $L1_2$ structure in the supercooled liquids using a map based on Steinhardt order parameters $(\hat{q_4},\hat{w_4})$ (details are provided in the supplemental material). The formation energies are given in Fig.~\ref{Fig3}. Very interestingly, the results show that the formation energies for $L1_2$ clusters are much lower than for the $L1_0$ cluster. This provides a first insight into the different behavior that takes place in the supercooled liquid for $x_{Cu}=0.25$ and $x_{Cu}=0.75$, which can give rise to $L1_2$ clusters, and for $x_{Cu}=0.5$, in which $L1_0$ is the ordered structure. Furthermore, the formation energy for $L1_2$ clusters is below $0.1eV$ for both $x_{Cu}=0.25$ and $x_{Cu}=0.75$, which is  lower than the formation energy of ISRO ($k=0$) and of the connected icosahedra MRO with the highest number density ($k \le 3$). This means that the formation of $L1_2$ clusters of $CuAu_3$ and $Cu_3Au$ is energetically favored. On the other hand, in the $x_{Cu}=0.5$ case, the formation energy for $L1_0$ clusters is close to $0.11eV$, comparable to that of connected icosahedra for $k \ge 1$. Thus, the formation of such $L1_0$ clusters is much less frequent than of $L1_2$ clusters (see Fig.~\ref{Fig3}).

\begin{figure}
\begin{center}
\includegraphics*[width=7cm]{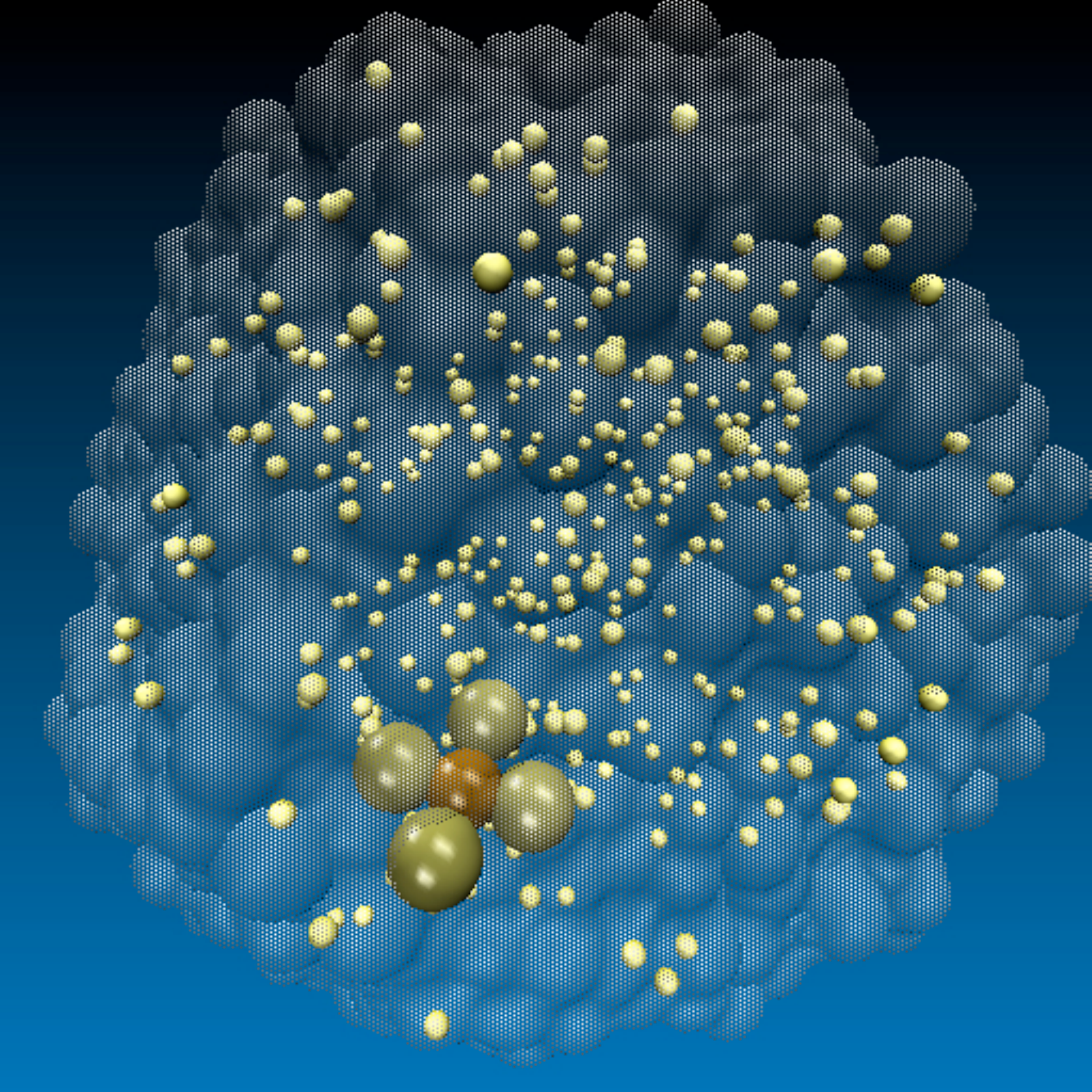}(a)
\includegraphics*[width=7cm]{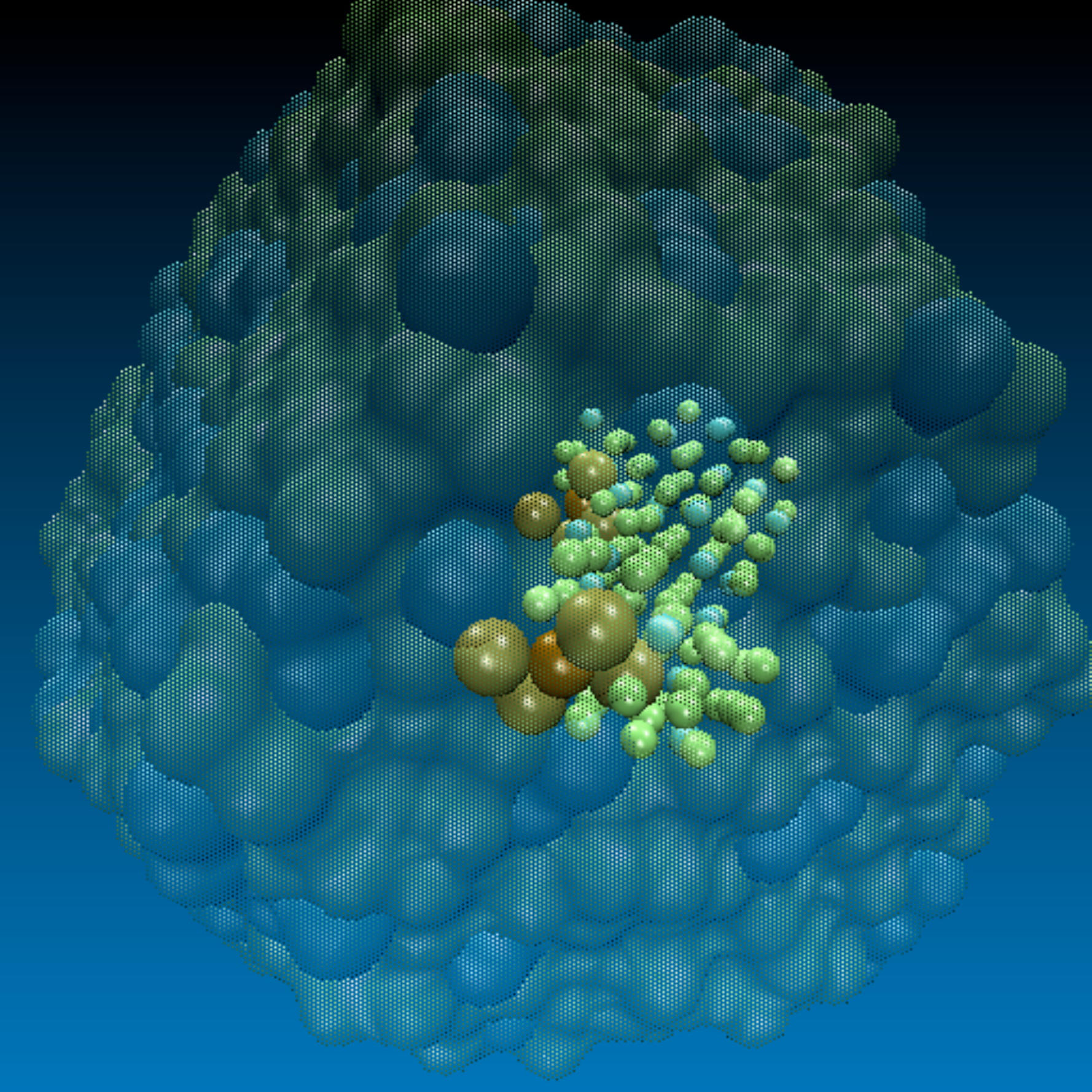}(b)
\end{center}
\caption{$L1_2$ ordered clusters of $Cu_3Au$ in the supercooled liquid (a) and in a configuration containing a critical nucleus (b). For clarity, only the central $Cu$ atom (ochre) and the four closest $Au$ (tan) are shown as large spheres and liquid-like atoms are translucent. In (a), atoms in icosahedra are shown as small yellow spheres. In (b), atoms from the critical nucleus are shown as small blue spheres ($Au$) and small green spheres ($Cu$).}
\label{Fig4}
\end{figure}

This provides the following picture. For $x_{Cu}=0.25$ and $x_{Cu}=0.75$, locally favored structures, that are both structurally and chemically ordered, easily form in the supercooled liquid. An example of a $L1_2$ cluster of $Cu_3Au$ is shown on the snapshot of Fig.~\ref{Fig4}(a). These clusters serve as precursors for the crystal nucleus, unlike icosahedral clusters in supercoooled liquid metals. The latter have a structure incompatible with the crystal, resulting in higher free energy barriers and in the remarkable stability of supercooled liquid metals. Here, the structural compatibility between $L1_2$ and the stable phase (the structurally ordered and chemically disordered solid solution) facilitates the nucleation process, resulting in the decrease in the free energy barrier observed in Fig.~\ref{Fig1}(a) for both $CuAu_3$ and $Cu_3Au$. This synergy continues throughout the nucleation step: the center of the nucleus converts into the stable phase, while $L1_2$ clusters continue to form on its surface (see snapshot of a critical nucleus of $Cu_3Au$ in Fig.~\ref{Fig4}(b)) and, in turn, assist crystallization. Such a mechanism is reminiscent of the role played by the body-centered cubic (BCC) phase during the nucleation of the stable face-centered cubic (FCC) phase from supercooled Lennard-Jones (LJ) liquids.\cite{oxtoby1998nucleation,tenWolde} Indeed, in the LJ case, nucleation starts with the formation of small BCC clusters that serve as precursors. Then, the center of the LJ nucleus converts into the FCC structure, while BCC clusters continue to form on its surface to promote crystallization. On the other hand, the higher formation energy for $L1_0$ clusters implies that the nucleation of the $CuAu$ is not promoted by ordered precursors and, as a result, the free energy barrier is much higher.

We finally compare the nucleation rates for the three systems. As discussed by Song {\it et al.}, the attachment rate ($f^{+}_{n_c}$) can, in some cases, be the dominant factor in the nucleation rate.\cite{song2018nucleation} Thus, following Frenkel and co-workers,\cite{Auer,auer2004quantitative,valeriani2005rate,filion2010crystal}, we determine $f^{+}_{n_c}$ for each system (see supplemental information for details). We find that $f^{+}_{n_c}$ are comparable for the three systems, varying from $2.6 \times 10^{-14}~s^{-1}$ ($x_{Cu}=0.25$) to $2.7 \times 10^{-14}~s^{-1}$ ($x_{Cu}=0.5$), and finally to $4.5 \times 10^{-14}~s^{-1}$ ($x_{Cu}=0.75$), and that the three systems exhibit kinetic prefactors of the same order. This means that the free energy barrier is the dominant factor when comparing the nucleation rates of the three alloys. Thus, given its greater free energy barrier, the nucleation rate for $CuAu$ is of the order of $10^{32}~m^{-3}.s^{-1}$ and is about four order of magnitude smaller than those obtained both for $CuAu_3$and $Cu_3Au$.

The analysis of the nucleation pathway and of its free energy profile allows us to identify a new nucleation mechanism, in which ordered precursors facilitate the formation of a critical nucleus of the solid solution. This result is linked to the presence in the parent phase of small ordered $L1_2$ clusters, that are energetically favored over competing low-energy structures like icosahedra. The $L1_2$ structure continues to promote nucleation later on as clusters keep forming at the solid-liquid interface during the development of the nucleus, resulting in a lower free energy barrier of nucleation. On the other hand, the $L1_0$ structure does not promote the nucleation process, leading us to connect the high formation energy of ordered $L1_0$ clusters to a higher free energy barrier of nucleation. As captured by Ostwald's step rule,\cite{ostwald1897studien} crystallization generally proceeds through stages that result in a gradual increase in order. Here, however, the mechanism observed involves an intermediate that is actually more organized than the end result, since it is both chemically and structurally ordered as summarized below 
\begin{center} 
$\begin{array}{ccccc}
\textrm{liquid} &\to&L1_2~\textrm{precursor}&\to&\textrm{solid solution}\\
\textrm{(structural disorder)}&&\textrm{(structural order)}&&\textrm{(structural order)}\\
\textrm{(chemical disorder)}&&\textrm{(chemical order)}&&\textrm{(chemical disorder)}\\
\end{array}$
\end{center} 
This novel pathway suggests new ways to understand and control the formation of solid solutions, as these have emerged recently as promising candidates for many applications including water splitting,\cite{maeda2005gan} hydrogen production\cite{ye2016high} and multi-principal elements and high entropy alloys.\cite{senkov2015accelerated,wu2016thermal}

\vspace{1cm}

{\bf Acknowledgements}
Partial funding for this research was provided by NSF through CAREER award DMR-1052808.\\


\end{document}